\documentclass{article}

\usepackage{arxiv}

\usepackage[utf8]{inputenc} 
\usepackage[T1]{fontenc}    
\usepackage{hyperref}       
\usepackage{url}            
\usepackage{booktabs}       
\usepackage{amsfonts}       
\usepackage{amssymb}        
\usepackage{amsmath}        
\usepackage{nicefrac}       
\usepackage{microtype}      
\usepackage{lipsum}		
\usepackage{graphicx}
\usepackage[numbers]{natbib}
\usepackage{doi}
\usepackage{wrapfig}

\title{The free algebra in R}

\author{ \href{https://orcid.org/0000-0001-5982-0415}{\includegraphics[width=0.03\textwidth]{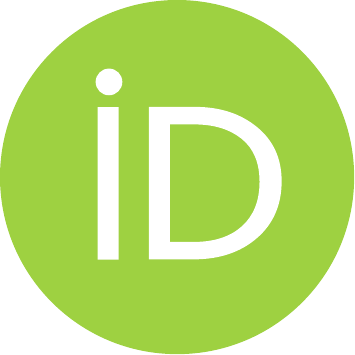}\hspace{1mm}Robin K. S.~Hankin}\thanks{\href{https://academics.aut.ac.nz/robin.hankin}{work};  
\href{https://www.youtube.com/watch?v=JzCX3FqDIOc&list=PL9_n3Tqzq9iWtgD8POJFdnVUCZ_zw6OiB&ab_channel=TrinTragulaGeneralRelativity}{play}} \\
 Auckland University of Technology\\
	\texttt{hankin.robin@gmail.com} \\
}

\renewcommand{\shorttitle}{The free algebra in R}

\hypersetup{
pdftitle={The free algebra in R},
pdfsubject={q-bio.NC, q-bio.QM},
pdfauthor={Robin K. S.~Hankin},
pdfkeywords={Free algebra}
}

\usepackage{Sweave}
\begin{document}
\maketitle

\begin{abstract}
  The free algebra is an interesting and useful algebraic object.
  Here I introduce {\tt freealg}, an R package which furnishes
  computational support for free algebras.  The package uses the
  standard template library's {\tt map} class for efficiency, which
  uses the fact that the order of the terms is algebraically
  immaterial.  The package follows {\tt disordR} discipline.
  I demonstrate some properties of free algebra using the package, and
  showcase package idiom.  The package is available on CRAN
  at \url{https://CRAN.R-project.org/package=freealg}.
\end{abstract}

\section{The free algebra}

\setlength{\intextsep}{0pt}
\begin{wrapfigure}{r}{0.2\textwidth}
  \begin{center}
\includegraphics[width=1in]{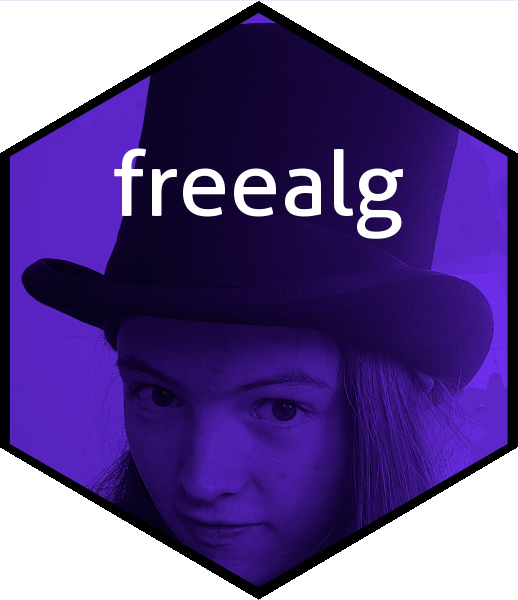}
  \end{center}
\end{wrapfigure}
The free algebra is the free R-module with a basis consisting of all
words over an alphabet of symbols with multiplication of words defined
as concatenation.  Such objects are a natural set to consider and have
a sum and product given by algebraic addition and string concatenation
respectively; the system is assumed to be distributive.  Formally, we
consider the free R-module with a basis consisting of all words over
an alphabet of symbols [conventionally lower-case letters] with
multiplication of words defined as concatenation.  The system inherits
associativity from associativity of concatenation; distributivity
follows from the definition of R-module.  However, the free algebra is
not commutative in general.  Thus, with an alphabet of $\{x,y,z\}$ and
$\alpha,\beta,\gamma,\delta\in\mathbb{R}$, we might define free
algebra elements $A,B$ as

\[
A=\alpha x^2yx + \beta zy\qquad B=\gamma z + \delta y^4
\]

we would then have

\begin{eqnarray*}
AB &=& \left(\alpha x^2yx+\beta zy\right)\left(\gamma z+\delta y^4\right)=\alpha\gamma x^2yxz+\alpha\delta x^2yxy^4+\beta\gamma zyz+\beta\delta zy^5\\
BA &=&\left(\gamma z+\delta y^4\right)\left(\alpha x^2yx+\beta zy\right)=\alpha\gamma zx^2yx + \alpha\delta y^4 x^2yx + \beta\gamma z^2y + \beta\delta y^4zy.
\end{eqnarray*}

The above examples are a little too general for the {\tt freealg}
package; the idiom requires that we have specific numerical values for
the coefficients $\alpha,\beta,\gamma,\delta$.  Here we will use
$1,2,-2,3$ respectively.

\begin{Schunk}
\begin{Sinput}
> library("freealg")
> (A <- as.freealg("xxyx + 2zy"))
\end{Sinput}
\begin{Soutput}
free algebra element algebraically equal to
+ 1*xxyx + 2*zy
\end{Soutput}
\begin{Sinput}
> (B <- as.freealg("-2z + 3yyyy"))
\end{Sinput}
\begin{Soutput}
free algebra element algebraically equal to
+ 3*yyyy - 2*z
\end{Soutput}
\end{Schunk}

Note that the terms are stored in an implementation-specific order.
For example, {\tt A} might appear as {\tt xxyz + 2*zy} or the
algebraically equivalent form {\tt 2*zy + xxyz} (see the {\tt disordR}
package~\cite{hankin2022_mvp_arxiv,hankin2022_disordR}).

\section{Computational implementation of free algebra: the {\tt STL}
map class}

A ``map'' is a sorted associative container that stores key-value
pairs with unique keys~\citep{musser2009}.  It is interesting here
because search and insertion operations have logarithmic complexity.
Free algebra objects are considered to be the sum of a finite number
of {\em words}, each multiplied by a coefficient.  A word is something
like $x^2yx$, represented internally as a list of signed integers:
usually, one identifies {\tt a} with 1, {\tt b} with 2, and so on, so
$x$ would be 24 and $x^2yz$ would be {\tt [24 24 25 24]}.  It is
understood that powers are nonzero.  An {\tt mvp} object is a map from
terms to their coefficients; thus $B= -2z + 3y^4$ might be

\begin{verbatim}
{[25,25,25,25]} -> 7, [26,25]} -> -2}
\end{verbatim}

We understand that coefficients are nonzero.  In {\tt C++} the
declarations would be

\begin{verbatim}
using namespace std;
using namespace Rcpp; 
typedef std::list<signed int> word; // a 'word' object is a list of signed ints
typedef map <word, double> freealg; // a 'freealg' maps word objects to reals
\end{verbatim}

Thus a {\tt word} is a list of signed {\tt int}s, and a {\tt freealg}
maps {\tt word} objects to doubles.  One reason why the {\tt map}
class is fast is that the order in which the keys are stored is
undefined: the compiler may store them in the order which it regards
as most propitious.  This is not an issue for the maps considered here
as addition and multiplication are commutative and associative.  The
package uses {\tt disordR} discipline~\cite{hankin2022_disordR}.  Note
also that constant terms are handled with no difficulty (constants are
simply maps from the empty map to its value), as is the zero
polynomial (which is simply an empty map).

\section{The package in use}

Free algebra objects have naturally defined addition and
multiplication, implemented by the package.  With $A,B$ as defined
above:

\begin{Schunk}
\begin{Sinput}
> A+B
\end{Sinput}
\begin{Soutput}
free algebra element algebraically equal to
+ 1*xxyx + 3*yyyy - 2*z + 2*zy
\end{Soutput}
\begin{Sinput}
> A*B
\end{Sinput}
\begin{Soutput}
free algebra element algebraically equal to
+ 3*xxyxyyyy - 2*xxyxz + 6*zyyyyy - 4*zyz
\end{Soutput}
\begin{Sinput}
> B*A
\end{Sinput}
\begin{Soutput}
free algebra element algebraically equal to
+ 3*yyyyxxyx + 6*yyyyzy - 2*zxxyx - 4*zzy
\end{Soutput}
\end{Schunk}

Note again that the terms are stored in an implementation-specific
order.  Inverses are coded using upper-case letters:

\begin{Schunk}
\begin{Sinput}
> A*as.freealg("X") # X = x^{-1}
\end{Sinput}
\begin{Soutput}
free algebra element algebraically equal to
+ 1*xxy + 2*zyX
\end{Soutput}
\end{Schunk}

See how multiplying by $X=x^{-1}$ on the right cancels one of the {\tt
  x} terms in {\tt A}.  Also note the transparent implementation of
(right) distributivity.  We can use this device in more complicated
examples:

\begin{Schunk}
\begin{Sinput}
> (C <- as.freealg("3 + 5X - 2Xyx"))
\end{Sinput}
\begin{Soutput}
free algebra element algebraically equal to
+ 3 + 5*X - 2*Xyx
\end{Soutput}
\begin{Sinput}
> A*C
\end{Sinput}
\begin{Soutput}
free algebra element algebraically equal to
+ 5*xxy + 3*xxyx - 2*xxyyx + 6*zy + 10*zyX - 4*zyXyx
\end{Soutput}
\begin{Sinput}
> C*A
\end{Sinput}
\begin{Soutput}
free algebra element algebraically equal to
- 2*Xyxxxyx - 4*Xyxzy + 10*Xzy + 3*xxyx + 5*xyx + 6*zy
\end{Soutput}
\end{Schunk}

With these objects we may verify that the distributive and associative
laws hold:

\begin{Schunk}
\begin{Sinput}
> c(A*(B+C) == A*B + A*C  ,  (A+B)*C == A*C + B*C  ,  A*(B*C) == (A*B)*C)
\end{Sinput}
\begin{Soutput}
[1] TRUE TRUE TRUE
\end{Soutput}
\end{Schunk}

\subsection{The commutator bracket and the Jacobi identity}

Various utilities are included in the package.  For example, the
commutator bracket is represented by reasonably concise idiom:

\begin{Schunk}
\begin{Sinput}
> a <- as.freealg("a")
> b <- as.freealg("b")
> .[a,b] # returns ab-ba
\end{Sinput}
\begin{Soutput}
free algebra element algebraically equal to
+ 1*ab - 1*ba
\end{Soutput}
\end{Schunk}

Using {\tt rfalg()} to generate random free algebra objects, we may
verify the Jacobi identity:

\begin{Schunk}
\begin{Sinput}
> x <- rfalg()
> y <- rfalg()
> z <- rfalg()
> .[x,.[y,z]] + .[y,.[z,x]] + .[z,.[x,y]]
\end{Sinput}
\begin{Soutput}
free algebra element algebraically equal to
0
\end{Soutput}
\end{Schunk}

\subsection{Substitution}

One of the advantages of working with the map class is that
substitution has a natural ready idiom:

\begin{Schunk}
\begin{Sinput}
> subs("aabccc",b="1+3x")  # aa(1+3x)ccc
\end{Sinput}
\begin{Soutput}
free algebra element algebraically equal to
+ 1*aaccc + 3*aaxccc
\end{Soutput}
\end{Schunk}

\begin{Schunk}
\begin{Sinput}
> subs("abccc",b="1+3x",x="1+d+2e")
\end{Sinput}
\begin{Soutput}
free algebra element algebraically equal to
+ 4*accc + 3*adccc + 6*aeccc
\end{Soutput}
\end{Schunk}

\subsection{Calculus}

There is even some experimental functionality for calculus:

\begin{Schunk}
\begin{Sinput}
> deriv(as.freealg("aaaxaa"),"a")
\end{Sinput}
\begin{Soutput}
free algebra element algebraically equal to
+ 1*aaaxa(da) + 1*aaax(da)a + 1*aa(da)xaa + 1*a(da)axaa + 1*(da)aaxaa
\end{Soutput}
\end{Schunk}

Above, {\tt da} means the differential of {\tt a}.  Note how it may
appear at any position in the product, not just the end (cf matrix
differentiation).

\section{Numerical verification: an example from matrix algebra}

With $A,B$ as above, we can verify that matrices, which obey all the
relations of free algebra, are consistent with the package:

\begin{Schunk}
\begin{Sinput}
> A
\end{Sinput}
\begin{Soutput}
free algebra element algebraically equal to
+ 1*xxyx + 2*zy
\end{Soutput}
\begin{Sinput}
> B
\end{Sinput}
\begin{Soutput}
free algebra element algebraically equal to
+ 3*yyyy - 2*z
\end{Soutput}
\begin{Sinput}
> A*B
\end{Sinput}
\begin{Soutput}
free algebra element algebraically equal to
+ 3*xxyxyyyy - 2*xxyxz + 6*zyyyyy - 4*zyz
\end{Soutput}
\end{Schunk}

Then we define three random matrices $x,y,z$, chosen to be $5\times
5$:

\begin{Schunk}
\begin{Sinput}
> x <- matrix(rnorm(25),5,5)
> y <- matrix(rnorm(25),5,5)
> z <- matrix(rnorm(25),5,5)
\end{Sinput}
\end{Schunk}

We may then translate the free algebra calculations into R matrix
idiom:

\begin{Schunk}
\begin{Sinput}
> A_matrix <- x %*% x %*% y %*% x + 2*z %*% y
> B_matrix <- 3*y %*% y %*% y %*% y -2*z
\end{Sinput}
\end{Schunk}

We then calculate the matrix product $AB$ in two ways:

\begin{Schunk}
\begin{Sinput}
> AB_matrix_way1 <- A_matrix %*% B_matrix
> AB_matrix_way2 <- (
+ +3*x %*% x %*% y %*% x %*% y %*% y %*% y %*% y
+ -2*x %*% x %*% y %*% x %*% z
+ +6*z %*% y %*% y %*% y %*% y %*% y
+ -4*z %*% y %*% z
+ )
\end{Sinput}
\end{Schunk}

Above we calculate the product firstly using R's matrix multiplication
{\tt \%*\%} and secondly using the {\tt freealg} product operation.
The two methods should agree:

\begin{Schunk}
\begin{Sinput}
> AB_matrix_way1
\end{Sinput}
\begin{Soutput}
           [,1]      [,2]      [,3]       [,4]       [,5]
[1,] -4467.0106 -3335.435 4128.9699 -3027.0180  614.49694
[2,]  -640.4003  -495.393  755.9870  -563.6747  -16.01638
[3,]  -639.9635  -426.448  314.8182  -263.9274  345.99543
[4,] -4976.4209 -3836.623 5209.0127 -3431.0073 -339.73926
[5,] -2745.9901 -2006.262 2229.6004 -1743.8982  749.56178
\end{Soutput}
\begin{Sinput}
> AB_matrix_way2
\end{Sinput}
\begin{Soutput}
           [,1]      [,2]      [,3]       [,4]       [,5]
[1,] -4467.0106 -3335.435 4128.9699 -3027.0180  614.49694
[2,]  -640.4003  -495.393  755.9870  -563.6747  -16.01638
[3,]  -639.9635  -426.448  314.8182  -263.9274  345.99543
[4,] -4976.4209 -3836.623 5209.0127 -3431.0073 -339.73926
[5,] -2745.9901 -2006.262 2229.6004 -1743.8982  749.56178
\end{Soutput}
\begin{Sinput}
> AB_matrix_way1 - AB_matrix_way2
\end{Sinput}
\begin{Soutput}
              [,1]          [,2]          [,3]          [,4]          [,5]
[1,]  9.094947e-13  4.547474e-13 -9.094947e-13  4.547474e-13  1.136868e-13
[2,] -3.410605e-13 -6.821210e-13  6.821210e-13  1.136868e-13  4.263256e-13
[3,]  3.410605e-13  4.547474e-13 -3.979039e-13 -2.842171e-13  0.000000e+00
[4,] -9.094947e-13  4.547474e-13  0.000000e+00  0.000000e+00 -1.136868e-13
[5,]  1.364242e-12  1.364242e-12 -2.273737e-12 -1.136868e-12  2.273737e-13
\end{Soutput}
\end{Schunk}

and indeed we see only small numerical rounding error.

\section{Conclusions and further work}

The {\tt freealg} package furnishes R-centric computational support
for working with the free algebra, leveraging the efficiency of the
{\tt STL} map class.  Further work might include utilities for
manipulating free Lie algebras.

\bibliographystyle{apalike}
\bibliography{freealg_arxiv}

\end{document}